\documentclass[aps,groupedaddress,a4paper,floatfix,twocolumn,footinbib,showpacs]{revtex4}
\usepackage{pslatex}
\usepackage{makeidx}
\usepackage{amsmath}
\usepackage{amsfonts}
\usepackage{amssymb}
\usepackage{mathrsfs}
\usepackage[mathscr]{euscript}
\usepackage[dvips]{graphicx}
\usepackage{fancyhdr}
\usepackage[hypertex=true]{hyperref}

\newcommand{\sech}{{\textrm{ sech}}}
\newcommand{\sgn}{{\textrm{ sgn}}}

\def\overstrike#1#2{{\setbox0\hbox{$#2$}\hbox to \wd0{\hss
    $#1$\hss}\kern-\wd0\box0}}

\begin{document}

\title{Few-cycle soliton propagation}
\author{P. Kinsler}
\affiliation{
  Department of Physics, Imperial College London,
  Prince Consort Road,
  London SW7 2BW, 
  United Kingdom.
}
\author{G.H.C. New}
\affiliation{
  Department of Physics, Imperial College London,
  Prince Consort Road,
  London SW7 2BW, 
  United Kingdom.
}

\date{\today}

\begin{abstract}

Soliton propagation is usually 
described in the ``slowly varying envelope approximation'' 
(SVEA) regime, which is not
applicable for ultrashort pulses.  
We present theoretical results and numerical simulations for 
both NLS and parametric ($\chi^{(2)}$) ultrashort solitons in the 
``generalised few-cycle envelope approximation'' (GFEA) regime, 
demonstrating their
altered propagation.

\end{abstract}


\maketitle

\lhead{
\includegraphics[height=5mm,angle=0]{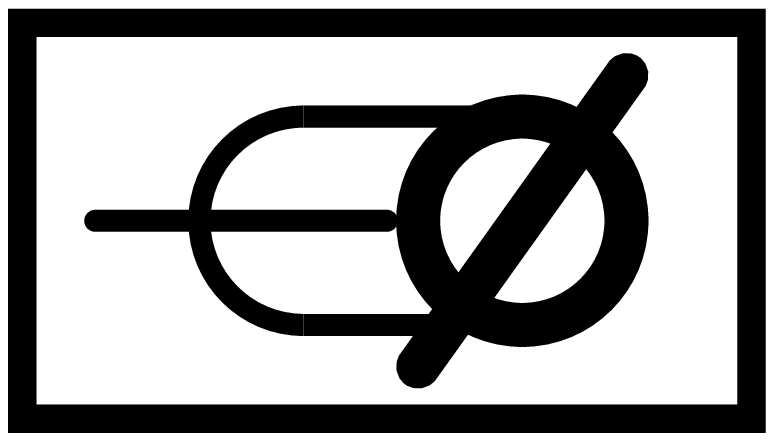}~~
FCSOLITON}
\chead{~}
\rhead{
\href{mailto:Dr.Paul.Kinsler@physics.org}{Dr.Paul.Kinsler@physics.org}\\
\href{http://www.kinsler.org/physics/}{http://www.kinsler.org/physics/}
}

\thispagestyle{fancy}

\noindent
{\em
NOTE: this work was published as 
Phys.Rev.A{\bf 69}, 013805 (2004).}

\section{Introduction}\label{S-intro}

This paper demonstrates the behavior of both Kerr and parametric solitons in
the few-cycle regime.  We compare  the usual ``many cycle'' or ``narrowband''
Slowly Varying Envelope Approximation''
(SVEA) descriptions to those obtained with the more accurate ``Generalised
Few-cycle Envelope Approximation'' (GFEA) theory.  We present a range of
results using both analytic and numerical methods.

There have been many suggestions for using solitons in communications as well
as optical logic (see e.g. 
\cite{Trillo-WS-1988ol,Potasek-1989jap,Drummond-BMS-1994spp}).  These
generally emphasise the stability of soliton profiles over long
propagation distances, and the potential for higher data rates with
shorter solitons 
\cite{Gromek-T-2000c,Zaspel-MRG-2001prb}.
It is clear that the
applicability of the nonlinear Schroedinger (NLS) equation to Kerr soliton
propagation is only valid in the SVEA
regime, where the soliton pulse contains many optical cycles.  However, since
the shorter the pulse, the fewer optical cycles it contains, the drive
to ultrashort solitons will eventually reach the 
few-cycle regime.

We discuss the change in propagation properties of few-cycle solitons when the
GFEA theory is used.  A description of the GFEA propagation equation is
presented in \cite{Kinsler-N-2003pra}, and a detailed derivation is given in
\cite{Kinsler-FCPP}.  It emerges that few-cycle pulses receive an extra
``phase twist'' compared with many-cycle pulses, and this raises the question
of how the fundamental characteristics of soliton propagation might be
preserved in the few-cycle regime.  

In simple cases, an SVEA propagation equation can be converted to the
corresponding GFEA form, accurate in the few-cycle limit, merely by 
applying the following operator to the polarisation term --
~
\begin{eqnarray}
    \frac{\left(1 + \imath \partial_t / \omega_0 \right)^2}
         {\left( 1 + \imath \sigma \partial_t / \omega_0 \right)}
&=&
  \left[
    1 
       + \imath 
           \left( 2 - \sigma \right)
           \frac{ \partial_t }{ \omega_0 }
       -  \left( 1 - \sigma \right)^2
           \frac{ \partial_t^2 }{ \omega_0^2  }
\right.
\nonumber
\\
&& ~~ \left.
       +  \imath \sigma 
          \left( 1 - \sigma \right)^2
           \frac{ \partial_t^3 }{ \omega_0^3 }
       + \mathscr{O}(\partial_t^4 / \omega_0^4 )
  \right]
\label{eqn-expandB2nd}
.
\end{eqnarray}

Here $\partial_t$ is used
as a compact notation for $d/dt$; $\omega_0$ is the carrier 
frequency, and $\sigma$ is the ratio of the group velocity to 
the phase velocity.

\section{NLS solitons}\label{S-nlss}

NLS solitons are hyperbolic secant (sech shaped) pulses that rely on the
interplay of third order $\chi^{(3)}$ (Kerr) nonlinearity and the material
dispersion to
propagate without changing.  It is relatively simple to derive the necessary
``nonlinear Schroedinger'' (NLS) equation for optical pulses in a Kerr
nonlinear medium in the SVEA limit.
As usual we write the field in the form  
$E(t,z) = A(t,z) e^{-\imath \omega_0 t} + A^*(t,z) e^{\imath \omega_0 t}$,
where the envelope $A$ varies slowly in comparison to the 
carrier period.

The lowest order (bright) soliton solution of the SVEA NLS equation is 
a hyperbolic secant pulse, which at $z=0$ will be 
~
\begin{eqnarray}
A(t,0) 
&=&
\eta \sech \left( \eta t \right).
\label{eqn-nlss-ic}
\end{eqnarray}

If instead we apply the more general GFEA to the 
propagation of optical pulses in a Kerr  medium,
we find that the pulse envelope of the soliton evolves 
according to:
~
\begin{eqnarray}
\partial_z
  A 
+ 
  \beta_1 \partial_t A 
+ 
  \frac{\imath \beta_2}{2}
  \partial_t^2 A 
&=&
\imath
\chi
    \frac{\left(1 + \imath \partial_t / \omega_0 \right)^2}
         {\left( 1 + \imath \sigma \partial_t / \omega_0 \right)}
\left| A \right|^2 A
\label{eqn-nlss-propagation}
\\
&\approx&
\imath
\chi
  \left[ 1 
        + 
         \left(2-\sigma\right)
         \imath \frac{\partial_t}{\omega_0}
  \right]
\left| A \right|^2 A
.
\label{eqn-nlss-propagation-approx}
\end{eqnarray}

We get the approximate form in eqn.(\ref{eqn-nlss-propagation-approx}) by 
truncating the expansion of the GFEA operator given in 
eqn.(\ref{eqn-expandB2nd}) to  first order in $\partial_t$ terms.
In the SVEA-equivalent limit where 
$\left| \partial_t A \right|/\omega_0 \rightarrow
0$ limit, the RHS nonlinear term is just $\imath \chi \left| A \right|^2 A$,
and the equation reduces to the standard NLS equation.  
The group velocity to phase velocity 
ratio is $\sigma = \omega_0 \beta_1 / \beta_0$, where $\beta_n
= \left. d^n k(\omega)/ d\omega^n \right|_{\omega_0}$.

Readers will notice that the mathematical form of the 
first order GFEA correction to eqn.(\ref{eqn-nlss-propagation})
appears similar to the so-called ``optical shock'' terms 
already discussed in the literature (e.g. 
see Zaspel\cite{Zaspel-1999prl,Park-H-2000prl},
or Potasek \cite{Potasek-1989jap}), which represents an intensity dependent 
group velocity.
A similar term also appears
in Biswas \& Aceves \cite{Biswas-A-2001jmo} who describe it as 
the ``self-steepening term for short pulses'' (see also 
\cite{Agrawal-NFO,Wabnitz-KA-1995oft}), which originates from a high-order 
dispersion effect, and is only indirectly a ``short pulse'' effect.
The origins for these terms are not the same as for the
GFEA correction terms, although they have the same self-steepening
effect.

Note the difference between our GFEA few-cycle terms and those 
given by the SEWA of Brabec and Krausz\cite{Brabec-K-1997prl}, 
which is due to their approximation of
$\sigma=1$. Close inspection of an alternative propagation equation 
valid for few-cycle pulses given by 
Trippenbach et. al. \cite{Trippenbach-WKBFB-2002oc} also reveals
a $\left( \sigma-2 \right)$ correction, present in the third nonlinear 
term on the RHS of 
their eqn.(14).

\subsection{GFEA Simulations}

The code developed for modelling the propagation of optical pulses in either
the SVEA or GFEA regimes is an improved version of the one used in [9]. 
Normalised units were based on a carrier frequency of $\omega_0 = 2 \pi \nu =
20\pi/\Delta $, unit nonlinear coefficient ($\chi=1$), pulse width ($\eta=1$),
and peak amplitude $A_0=1$.  In this scheme, $\Delta=1$ and $\Delta=3$
solitons contain about 10 cycles and 3 cycles respectively.  Distances
are normalised by the dispersion distance, so 
$\xi = z \left( \eta/\beta_2 \right)^{-1}$.
Our simulations implement the full GFEA correction to the
        polarization, as shown on the LHS of eqn.(\ref{eqn-expandB2nd}).

\begin{figure}[htbp]
\includegraphics[width=40mm,angle=-90]{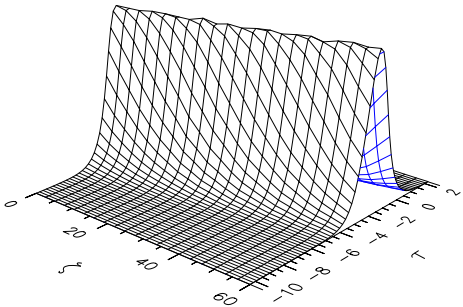}
\caption{
\label{F-soliton} 
GFEA soliton propagation. This shows the pump field 
profile (i.e. $\left| A \right|^2$)
for a $\Delta=3$ pulse with group/phase velocity ratio of $\sigma=1$.  The 
pulse starts with an offset of $\tau_0 = -5$.
}
\end{figure}

A typical result for single soliton propagation is shown in fig.
\ref{F-soliton}. Whereas under the SVEA the pulse would be stationary within 
the
display, there is now a drift to larger $t$ values (reduce group 
velocity) arising from few-cycle effects.   The results of simulations
over a range of values of $\Delta$ and $\sigma$ are shown in fig. 2.  The
presence of the $(2 - \sigma)$ term in eqn.(\ref{eqn-expandB2nd}) means that,
to first order, the few-cycle correction vanishes near $\sigma = 2$ and the
sign of the drift reverses at this point.    Moreover, as the bandwidth
represented by $\Delta$ gets larger (and the number of cycles correspondingly
fewer), the velocity change becomes more pronounced.  Note also that, for
$\sigma = 0$, only two correction terms appear in eqn.(\ref{eqn-expandB2nd}).

\begin{figure}[htbp]
\includegraphics[width=40mm,angle=-90]{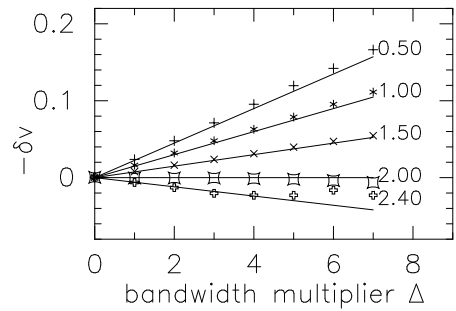}
\caption{
\label{F-soliton-dv} 
GFEA soliton velocity shifts  ($-\delta v$).
as a function of bandwidth multiplier $\Delta$.
The pulses at $\Delta=1$ are approximately 10 cycles long.
The symbols represent the velocity shifts obtained by 
numerical simulation, the solid lines the corresponding 
theoretical prediction.  
The numerical labels give the 
group/phase velocity ratio $\sigma$ corresponding to the 
nearest data points.
}
\end{figure}

\subsection{Theoretical Group Velocity Shift}

It is not necessary to rely on computer simulations to predict this few-cycle
shift in the group velocity, at least in the case of weak few-cycle effects. 
Biswas \& Aceves \cite{Biswas-A-2001jmo} have already provided a multi-scale
method giving the effects of various perturbations on standard NLS soliton
propagation.  
Their eqn.(13) applied to our NLS equation gives the velocity shift as
~
\begin{eqnarray}
v_A
&=&
\frac{d}{dz}
\frac{
  \int_{-\infty}^{+\infty} 
    t A^* A
    ~ dt
}
{ \int_{-\infty}^{+\infty} 
    A^* A
    ~ dt
}
,
\end{eqnarray}
~
which we can evaluate by inserting the SVEA
soliton profile from eqn.(\ref{eqn-nlss-ic})
into the few-cycle perturbation
term in
eqn.(\ref{eqn-nlss-propagation-approx}), namely
~
\begin{eqnarray}
R
&=&
 \chi
 \left( 2-\sigma \right) 
          \partial_t A^*A^2 / \omega_0
.
\label{eqn-nlss-perturbation}
\end{eqnarray}

This is of the same form as the $\lambda$ ``self-steepening'' term 
$\lambda \partial_t \left| q \right|^2 q $ in 
Biswas \& Aceves \cite{Biswas-A-2001jmo}, allowing for the changed 
notation and different prefactors.
Solving to first order in the perturbation 
%$\epsilon R$ (i.e. for $\epsilon \ll 1$)
and using the intermediate quantity 
~
$\bar{\tau}
=
  \int_{-\infty}^{+\infty} 
    t \partial_z \left( A^*A \right)
    ~ dt$
gives 
~
\begin{eqnarray}
\bar{\tau}
&=&
-
2 \chi
\frac{2-\sigma}{\omega_0}
  \eta^4
\frac{\pi}{2 \eta}
,
\\
v_A
&=&
-
\chi
\frac{2-\sigma}{\omega_0}
  \eta^2
.
\label{eqn-nlss-gfea-lambdalike}
\end{eqnarray}

With our parameters, 
this gives a velocity shift of
~
\begin{eqnarray}
\delta v &=& v_A = 
- 
\frac{ \left( 2-\sigma \right) \Delta }
     {20\pi}
\approx 
  - 0.01592 \left( 2-\sigma \right) \Delta
.
\label{eqn-nlss-dv}
\end{eqnarray}

Since we assume a small perturbation (eqn.(\ref{eqn-nlss-perturbation})),
it is clear that this prediction is most valid for higher carrier 
frequencies and/or
weaker nonlinearities, i.e. where the effect of the nonlinearity 
is small over the time of an optical cycle.

The predictions of eqn.(\ref{eqn-nlss-dv}) are plotted as solid lines in fig.
\ref{F-soliton-dv}, and the agreement with the numerical simulations 
is seen to be remarkably good, even for
very wideband pulses (e.g. the $\sim$1 cycle cases
where $\Delta=8$). However, the effect of the higher order 
GFEA contributions, present in the simulations but not 
in eqn.(\ref{eqn-nlss-dv}), becomes visible near $\sigma=2$, where the first 
order corrections become small.

\section{Parametric Solitons}\label{S-psol}

Parametric solitons, otherwise known as ``simultons'' or 
``quadratic solitons'',  rely on the interplay between dispersion
and a {\it second order} $\chi^{(2)}$ interaction to maintain fixed 
envelope profiles between a pair of pulses
propagating in tandem.

If we modify the standard (dimensionless) propagation equation of 
Werner \& Drummond\cite{Werner-D-1993josab} to include few-cycle
terms to first order, with carrier frequencies $\omega_\psi = 2 \omega_\phi$
and wavevectors $\beta_{0\psi}= 2 \beta_{0\phi}$, we get
~
\begin{eqnarray}
\partial_\xi
  \psi
+ 
  \frac{\imath \beta_{2\psi}}{ 2 \left| \beta_{2\phi} \right|}
  \partial_\tau^2 \psi
&=&
-
\left[
  1 + 
  \imath \frac{2-\sigma}{\omega_\psi}
  \partial_\tau
\right]
\frac{\phi^2}{2}
,
\\
\partial_\xi
  \phi
+ 
  \frac{\imath}{2}
  \sgn(\beta_{2\phi})
  \partial_\tau^2 \phi
&=&
\left[
  1 + 
  \imath \frac{2-\sigma}{\omega_\phi}
  \partial_\tau
\right]
 \psi
 \phi^*
.
\label{eqn-psol-propagdimless}
\end{eqnarray}

Here $\beta_{1\psi}, \beta_{1\phi}$ and $\beta_{2\psi}, \beta_{2\phi}$ 
are the group velocities and group velocity dispersions
respectively.  We work in a co-moving frame where $\beta_{1\psi} = 
\beta_{1\phi}$, so the two 
pulses remain co-propagating.  This ensures that their group-phase 
velocity ratios are identical ($\sigma = \sigma_\psi = \sigma_\phi$).
Distance and times are normalised using
$z_0^{-1} = \left| \chi \Phi_0 \right| = 
\left| \beta_{2\phi} \right| / t_0^2 $, and  $\xi = z / z_0$, $\tau = t / t_0$.

In the SVEA limit ($\left| \partial_\tau \psi \right|/\omega_\psi \rightarrow
0$, and $\left| \partial_\tau \phi \right|/\omega_\phi \rightarrow 0$), the
standard anzatz gives the solutions 
~
\begin{eqnarray}
  \phi(z,t) 
=
 \Phi / \Psi_0 
&=&
  \phi_0
  \sech^2(\kappa \tau) 
  \exp( \imath \theta_\phi \xi )
,
\label{eqn-psol-sveaphi}
%\nonumber
\\
  \psi(z,t) 
=
 \Psi / \Psi_0 
&=&
  \psi_0
  \sech^2(\kappa \tau) \exp( \imath \theta_\psi \xi )
,
\label{eqn-psol-sveapsi}
\end{eqnarray}
~
where $\theta_\phi = 2 \kappa^2 \sgn(\beta_{2\phi})$, $\theta_\psi + 2
\kappa^2 \left( \beta_{2\psi} / \left| \beta_{2\phi} \right| \right) -z_0
\left( \beta_{2\psi} - 2 \beta_{2\phi} \right)= 0$ are time independent
constants.  In our chosen frame of reference, these parametric soliton pulses
have a group velocity of zero in the SVEA limit.

\subsection{GFEA Simulations}

We use the same basic code as for the Kerr soliton simulations,
with the different form of nonlinearity.  We use the normalised
units described after eqn.(\ref{eqn-psol-propagdimless}).

\begin{figure}[htbp]
\includegraphics[width=40mm,angle=-90]{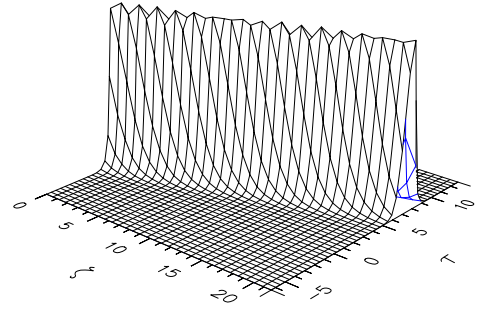}
\caption{
\label{F-fcsimulton} 
GFEA simulated simulton propagation: the pump field profile 
$\left| \psi \right|^2$ for a 3-cycle pulse ($\Delta=3$) with $\sigma=1$.
The signal field profile $\left| \phi \right|^2$ is similar.
}
\end{figure}

\begin{figure}[htbp]
\includegraphics[width=40mm,angle=-90]{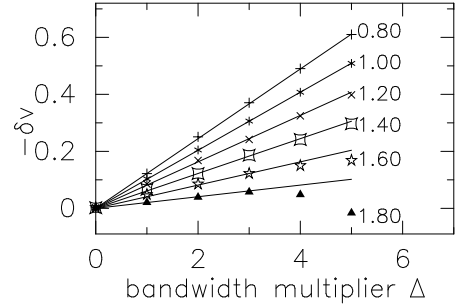}
\caption{
\label{F-fcsimulton-dv} 
GFEA simulated simulton velocity shifts $-\delta v$
as a function of bandwidth multiplier $\Delta$.
The pulses at $\Delta=1$ are approximately 10 cycles long.
The symbols represent the velocity shifts obtained by 
numerical simulation, the solid lines the corresponding 
theoretical prediction.
The numerical labels give the 
group/phase velocity ratio $\sigma$  corresponding to the 
nearest data points.
}
\end{figure}

Fig. \ref{F-fcsimulton} shows a typical result for the propagation of a
parametric soliton.  As for the Kerr soliton, few-cycle effects modify the
group velocity.  Results for different soliton widths $\Delta$ and different
$\sigma$ values are summarised in fig. \ref{F-fcsimulton-dv}.  The velocity
shift remains remarkably linear even for extremely wideband pulses.  Note that
$\Delta = 5$ corresponds to a 2 cycle pulse, for which the electric field
profile would not appear particularly $\sech^2$ shaped, because of the small
number of carrier oscillations.  In reality, however, the parabolic form of
the material dispersion assumed in the simulations will not be maintained over
such a wide bandwidth, and other distortions are likely to predominate over
few-cycle effects in these circumstances.   In addition, as the bandwidth
increases, the spectra of the two pulses will eventually overlap, despite the
separation  of their carrier frequencies.

\subsection{Theoretical Group Velocity Shift}

As described in section \ref{S-nlss}, we followed the method of 
Biswas and Aceves\cite{Biswas-A-2001jmo} to show 
how few-cycle corrections for NLS 
solitons would modify the group velocity.  We now show that the
same method can predict parametric soliton group velocity
shifts. Eqn.(13) from  Biswas and Aceves\cite{Biswas-A-2001jmo}
applied to the equations for parametric solitons gives
the velocity shifts as
~
\begin{eqnarray}
v_\phi
&=&
\frac{d}{d\xi}
\frac{
  \int_{-\infty}^{+\infty} 
    \tau \phi^*\phi
    ~ d\tau
}
{ \int_{-\infty}^{+\infty} 
    \phi^*\phi
    ~ d\tau
}, 
\label{eqn-psol-vphi}
\\
v_\psi
&=&
\frac{d}{d\xi}
\frac{
  \int_{-\infty}^{+\infty} 
    \tau \psi^*\psi
    ~ d\tau
}
{ \int_{-\infty}^{+\infty} 
    \psi^*\psi
    ~ d\tau
}
\label{eqn-psol-vpsi}
,
\end{eqnarray}
~
which we can evaluate by inserting the SVEA
soliton profiles 
(from eqn.(\ref{eqn-psol-sveaphi}) and (\ref{eqn-psol-sveapsi}))
along with the few-cycle perturbations
to the propagation:
~
\begin{eqnarray}
R_\phi 
&=&
 \left( 2-\sigma \right) 
          \partial_\tau \psi \phi^* / \omega_\psi
, 
\label{eqn-psol-perturbationphi}
\\
R_\psi
&=&
 -\left( 2-\sigma \right) 
          \partial_\tau \phi^2 / \left(2 \omega_\psi\right)
.
\label{eqn-psol-perturbationpsi}
\end{eqnarray}

Solving to first order in the perturbation(s) 
and using the intermediate quantity 
~
$\bar{\tau}_\phi
=
  \int_{-\infty}^{+\infty} 
    \tau \partial_\xi \left( \phi^*\phi \right)
    ~ d\tau$
gives 
~
\begin{eqnarray}
\bar{\tau}_\phi
&=&
\frac{2-\sigma}{\omega_\phi}
\left(
  \psi_0 \phi_0^* \phi_0
 +
  \psi_0^* \phi_0 \phi_0^*
\right)
\frac{-16\pi}{45}
,
\label{eqn-psol-taubarphi}
\\
v_\phi
&=&
-
\frac{8}{15}
\frac{2-\sigma}{\omega_\phi}
\left(
  \psi_0
 +
  \psi_0^*
\right)
,
\label{eqn-psol-vphi-value}
\end{eqnarray}
~
and similarly using 
~
$\bar{\tau}_\psi
=
  \int_{-\infty}^{+\infty} 
    \tau \partial_\xi \left( \psi^*\psi \right)
    ~ d\tau$,
~
\begin{eqnarray}
\bar{\tau}_\psi
&=&
-\frac{1}{2}
\frac{2-\sigma}{\omega_\psi}
\left(
  {\phi_0^*}^2 \psi_0
 +
  \phi_0^2 \psi_0^*
\right)
\frac{-16\pi}{45}
\label{eqn-psol-taubarpsi}
,
\\
v_\psi 
&=&
\frac{4}{15}
\frac{2-\sigma}{\omega_\psi}
\left(
  {\phi_0^*}^2 / \psi_0^*
 +
  \phi_0^2 / \psi_0
\right)
.
\label{eqn-psol-vpsi-value}
\end{eqnarray}

In our numerical simulations we used $\psi_0=3$, $\phi_0=-6\imath$, and
$\omega_\psi=2\omega_\phi=20\pi/\Delta$, so that 
\begin{eqnarray}
\delta v &=& v_\psi ~=~  v_\phi
\approx
-
0.1019 \left(2-\sigma\right) \Delta
.
\label{eqn-psol-dv}
\end{eqnarray}

Because the velocity
shift for each pulse of the pair making up the parametric soliton
is the same (at least for first order GFEA corrections), the 
two pulses remain co-propagating, and the soliton survives.

The predictions of eqn.(\ref{eqn-psol-dv}) are plotted as solid lines in fig.
\ref{F-fcsimulton-dv}, and the agreement with the numerical simulations 
is seen to be remarkably good, even for
wideband pulses (e.g.  $\Delta \ge 4$). However, the effect of the higher order 
GFEA contributions, present in the simulations but not 
in eqn.(\ref{eqn-psol-dv}), start to become
visible above $\sigma \approx 1.60$, since near $\sigma=2$
the first order few-cycle correction does not dominate.

\section{Conclusion}\label{S-conc}

We have investigated two types of soliton propagation beyond the standard
SVEA regime both theoretically and numerically.  
The most important result is that, according to the GFEA theory, 
soliton propagation remains robust in the few-cycle regime.  This is obviously 
encouraging for proposed applications involving ultrashort solitons -- 
although for such wideband pulses, there are other complications beyond
just the few-cycle ones examined in this paper.

The major effect of the shortening pulses is a 
group velocity shift, despite the fact that the perturbation term does
not look like a 
a simple group-velocity term.  It is likely that the few-cycle ``phase
twist'' added to the propagation also affects the other properties of 
soliton pulses, e.g. collisions, which has obvious potential 
implications for soliton-based ultrafast optical logic gates.

%\bibliography{../_pulse-fewcyc/fewcyc}

\end{document}